# Thin film growth by pulsed laser deposition and properties of 122-type iron-based superconductor $AE$(Fe$_{1-x}$Co$_x$)$_2$As$_2$ ($AE$ = alkaline earth)

Takayoshi Katase[1, *, †], Hidenori Hiramatsu[1], Toshio Kamiya[1], and Hideo Hosono[1, 2]

1: Materials and Structures Laboratory, Tokyo Institute of Technology, Mailbox R3-1, 4259 Nagatsuta-cho, Midori-ku, Yokohama 226-8503, Japan

2: Frontier Research Center, Tokyo Institute of Technology, S2-6F East, Mailbox S2-13, 4259 Nagatsuta-cho, Midori-ku, Yokohama 226-8503, Japan

($^*$) Corresponding author: katase@lucid.msl.titech.ac.jp

$^†$Present address: Frontier Research Center, Tokyo Institute of Technology, Yokohama, Japan

## Abstract

This paper reports comprehensive results on thin-film growth of 122-type iron-pnictide superconductors, $AE(\text{Fe}_{1-x}\text{Co}_x)_2\text{As}_2$ ($AE$ = Ca, Sr, and Ba, $AE\text{Fe}_2\text{As}_2$:Co) by a pulsed laser deposition method using a neodymium-doped yttrium aluminum garnet laser as an excitation source. The most critical parameter to produce the $\text{SrFe}_2\text{As}_2$:Co and $\text{BaFe}_2\text{As}_2$:Co phases is the substrate temperature ($T_s$). It is difficult to produce highly-pure $\text{CaFe}_2\text{As}_2$:Co phase thin film at any $T_s$. For $\text{BaFe}_2\text{As}_2$:Co epitaxial films, controlling $T_s$ at 800–850 °C and growth rate to 2.8–3.3 Å/s produced high-quality films with good crystallinity, flat surfaces, and high critical current densities > 1 MA/cm$^2$, which were obtained for film thicknesses from 100 to 500 nm. The doping concentration $x$ was optimized for $\text{Ba}(\text{Fe}_{1-x}\text{Co}_x)_2\text{As}_2$ epitaxial films, leading to the highest critical temperature of 25.5 K in the epitaxial films with the nominal $x = 0.075$.

## 1. Introduction

Iron-based superconductors have been studied intensively since the discovery of LaFe$Pn$O ($Pn$ = P, As) [1,2], and their critical temperatures ($T_c$) have reached 55 K [3], the highest value next to those of cuprates. Diverse studies have been devoted to revealing relationships between superconductivity and magnetism from experimental and theoretical viewpoints [4] because, similar to the cuprates, parent phases of iron-based superconductors exhibit anti-ferromagnetism and superconductivity emerges with its disappearance upon carrier doping. Although most of these works have been performed on polycrystalline samples, high-quality single crystals or epitaxial films are necessary to reveal intrinsic physical properties as well as for practical applications such as Josephson junction devices and high-field magnets. Thin film growth of iron-based superconductors has been examined since 2008 [5]. Until now, thin films exhibiting $T_c \geq$ 20 K have been reported for 1111-type $RE$FeAsO ($RE$ = La [6], Nd [7,8], and Sm [9]), 122-type $AE$Fe$_2$As$_2$ ($AE$ = Sr [10–12] and Ba [12–16]) and 11-type Fe(Se,Te) [17] systems. In particular, 122-type $AE$(Fe$_{1-x}$Co$_x$)$_2$As$_2$ ($AE$Fe$_2$As$_2$:Co) has been studied extensively because it is easier to grow as a superconducting epitaxial film than other systems due to the lower vapor pressure of the Co dopant compared with those of F and K dopants [10]. In addition, higher $T_c$ than those of the Co-doped 1111-type compounds have been obtained for this system although the Co are doped to the Fe site in the active pairing layers (i.e., direct doping) [18–21]. Among $AE$Fe$_2$As$_2$:Co, chemically stable BaFe$_2$As$_2$:Co (i.e., $AE$ = Ba) epitaxial films [13] have been examined most intensively and high critical current densities ($J_c$) $\geq$ 1 MA/cm$^2$ have been achieved [14,22,23]. Such progress has led to demonstrations of Josephson junctions [24,25] and a superconducting quantum interference device [26] based on the BaFe$_2$As$_2$:Co epitaxial films. However, their device performances were inferior to those of cuprates owing to the metallic character of the parent phase [24,26,27].

In contrast to the Josephson-device properties, an advantageous grain boundary property of BaFe$_2$As$_2$:Co has been clarified [27]; i.e., the intergrain $J_c$ of BaFe$_2$As$_2$:Co at [001]-tilt grain boundaries shows a transition from a strong link to a weak link at a critical misorientation angle of 9 degrees, which is much larger than those reported for cuprates (3–5 degrees [28]). This characteristic grain boundary property, along with high upper critical magnetic fields ($H_{c2}$) $\geq$ 50 T and small anisotropic factors $\gamma = H_{c2}//ab$ / $H_{c2}//c$ = 1–2 [29], is promising for producing wires [30,31] and tapes [32,33] exhibiting high-density supercurrents under high magnetic fields. The cuprates have high $J_c$ > 1 MA/cm$^2$ at 77 K but large $\gamma$ of up to 100 (the minimum $\gamma$ of YBa$_2$Cu$_3$O$_{7-\delta}$ (YBCO) is as large as 7.) [34], and other representative superconductors such as Nb$_3$Sn and MgB$_2$ have relatively low $H_{c2}$ < 30 T [35,36].

To date, unique techniques using buffer layers of Fe metal [37] and perovskite-type oxides such as SrTiO$_3$ and BaTiO$_3$ [14] have been proposed by two groups to produce high-quality BaFe$_2$As$_2$:Co films exhibiting high $J_c$. On the other hand, we succeeded in growing high-quality BaFe$_2$As$_2$:Co epitaxial films with $J_c$ > 1 MA/cm$^2$ directly on single-crystal insulating-oxide substrates without any buffer layers simply by optimizing growth conditions [22] using pulsed laser deposition (PLD). The above-mentioned groups that employ the buffer layer technique use KrF excimer lasers as the excitation sources in their PLD, while we employ a neodymium-doped yttrium aluminum garnet (Nd:YAG) laser [38]. The origin of this discrepancy in results is not yet clear because these PLD processes contain many differences such as the geometrical configuration of the growth chambers, the base pressure and vacuum system, the preparation process, and the quality of the targets as well as the excitation source. Systematic optimization of a film growth process is important in general because film quality depends strongly on the growth parameters. Therefore, we expect that systematic investigation of growth parameters may lead to the revealing of the above unclarified issues. However, there has been no report on this subject for the 122-type $AE$Fe$_2$As$_2$:Co system, notwithstanding that 122-type films are the most promising for thin-film tape applications among the iron-based superconductors. In this paper, we studied effects of growth parameters on structure, quality and physical properties of $AE$Fe$_2$As$_2$:Co ($AE$ = Ca, Sr, and Ba) films grown by PLD using a Nd:YAG laser.

## 2. Experimental

### 2.1. Thin film growth

Thin films were grown on (001)-oriented mixed-perovskite type (La,Sr)(Al,Ta)O$_3$ (LSAT, $a$/2 = 0.387 nm [39], where $a$ denotes the $a$-axis lattice parameter) and rock-salt type MgO ($a$ = 0.421 nm) single-crystal substrates by PLD using the second harmonic (wavelength: 532 nm) of a Nd:YAG laser (Spectra Physics, Model: INDI-40, pulse width < 10 nm, repetition rate 10 Hz) controlled with a digital pulse delay generator (Model: DG 535, Stanford Research System) (Fig. 1(a)). Growth rate was varied from 1.6 to 4.3 Å/s by changing the laser-pulse fluences from 1.7 to 4.3 J/cm$^2$/pulse. This change was performed by tuning the flash lamp voltage and/or the timing of the Q-switch trigger for the second harmonic generation. Typical film thicknesses were 300–400 nm, but were varied from 0.1 to 1 μm for the BaFe$_2$As$_2$:Co films in order to examine their thickness ($t$) dependence. The distance between the substrate and the PLD target disk was 30 mm. The base pressure of the PLD growth chamber was ca. $5 \times 10^{-7}$ Pa, but the vacuum became worse during the film deposition (e.g., up to $1 \times 10^{-5}$ Pa at substrate temperature ($T_s$) = 800 °C) due to outgas from the chamber surfaces.

The PLD targets (size: 6 mm × 9 mm × 5 mm thick, shown in Fig. 1(a)) were synthesized by solid-state reactions of stoichiometric mixtures of the intermediate compounds [$AE$As, Fe$_2$As, and Co$_2$As were synthesized from individual elemental reagents, $AE$ = Ca, Sr, and Ba (distilled, dendritic pieces, 99.99 % purity, Sigma-Aldrich), Fe (powders, > 99.9 % purity, Kojundo Chemical Laboratory), Co (powders, 99.9 % purity, Rare Metallic), and As (grains, 99.9999% purity, Kojundo Chemical Laboratory)] via reactions $AE$As + (1–$x$) Fe$_2$As + $x$ Co$_2$As → $AE$(Fe$_{1-x}$Co$_x$)$_2$As$_2$ [22]. In addition to the targets prepared for determination of the chemical composition ($x$) dependence of Ba(Fe$_{1-x}$Co$_x$)$_2$As$_2$, we prepared PLD targets for Ca(Fe$_{0.97}$Co$_{0.03}$)$_2$As$_2$, Sr(Fe$_{0.9}$Co$_{0.1}$)$_2$As$_2$, and Ba(Fe$_{0.92}$Co$_{0.08}$)$_2$As$_2$ because these $x$ values were reported as the optimal doping concentrations in refs. [19,21,40]. As reported in ref. [22], when preparing targets, fine cutting of the metallic $AE$ reagents is most important to obtain high-purity PLD targets because bulky $AE$ reagents easily remain unreacted in the resulting target disks. Therefore, it is necessary to synthesize well-reacted $AE$As intermediates prior to the final reactions. The mixtures of the intermediates were pressed into disks and then heated in evacuated silica-glass ampules at various reaction temperatures for 16 hours. All target preparation processes, except for the sealing of the pressed disks in glass ampules and the heating of the ampules, were performed in a glove box with a dry and inert atmosphere (dew point < –90 °C, oxygen concentration < 1ppm). The densities of the resulting sintered target disks were 60–70 %.

The single-crystal substrates were annealed at 1050 °C for 30 min in air to remove contaminants and improve surface morphology prior to the film deposition. The annealed substrate was tightly clamped by a top cover with a ~10 mm × 10 mm opening and a substrate carrier plate made from stainless steel (Figs. 1(b) and (c)). The back side of the substrate carrier was heated by halogen lamp irradiation (maximum power: 1.5 kW) focused with a gold mirror (Fig. 1(a)). $T_s$ was varied from 600 to 1000 °C, monitored with both a K-type (chromel / almel) thermocouple buried near the substrate carrier and a radiation thermometer. The monitored $T_s$ was calibrated using a thermocouple contacted to the substrate surface in preliminarily experiments. Such indirect substrate heating, along with good thermal contact (by inserting a thin platinum sheet between the substrate carrier and the substrate), is important to obtain homogeneous temperature distribution over the substrate surface (Fig. 1(d)). After deposition the substrate heater was turned off immediately. The substrate carrier was transferred to the preparation/load-lock chamber when $T_s$ became < 200 °C, and then was further cooled down to room temperature for 15–30 min.

### 2.2. Characterization

Crystalline phases of the target disks and films were determined by conventional X-ray diffraction (XRD, radiation: Cu K$\alpha_1$ and K$\alpha_2$, $\theta$–$2\theta$ synchronous scan). Although the angle resolution of conventional XRD is not sufficient to characterize highly-oriented epitaxial films, it is appropriate for detecting a small amount of impurity and non-oriented crystallites because of a stronger diffraction intensity and higher sensitivity than those of a high-resolution XRD. Fluctuations of crystallite orientation were characterized using rocking curves of out-of-plane ($2\theta$-fixed $\omega$ scans) and in-plane ($2\theta_\chi$-fixed $\phi$ scans) diffractions with a high-resolution XRD apparatus (radiation: Cu Kα$_1$, monochromated by Ge (220)). The film thicknesses were obtained with a stylus surface profiler by measuring the step height between the film-deposited and undeposited (formed at the substrate edges by the top clamping cover as shown in Fig. 1(c)) regions. Surface morphology of the films was observed by atomic force microscopy (AFM). The chemical composition of the resulting bulks and films were evaluated by X-ray fluorescence (XRF) analysis. XRD, film thickness, AFM, and XRF measurements were performed in ambient atmosphere at room temperature.

Transport $J_c$ ($J_c^{trans}$) at 4 K were determined with a criterion of 1 μV/cm from current density – voltage ($J$–$V$) curves using 300-μm-long and 8-μm-wide micro-bridges, which were patterned by photolithography and Ar ion milling. The magnetic $J_c$ ($J_c^{mag}$) at 3.5 K were extracted based on the Bean model [41] from magnetization hysteresis loops measured with a vibrating sample magnetometer. External magnetic fields ($H$) were applied parallel to the *c*-axis of the films. To estimate $x$ in BaFe$_2$As$_2$:Co films, temperature ($T$) dependences of in-plane electrical resistivity ($\rho$) were measured by the four-probe method under an applied dc current of 20 μA parallel to the *ab* plane of the film in the $T$ range of 2–300 K with a physical property measurement system, because the $T_c$ of BaFe$_2$As$_2$:Co clearly depends on the doping level $x$ [42].

## 3. Results and discussion

### 3.1. Phase stability of bulk samples

First, we confirmed that polycrystalline bulk samples were obtained for all the *AE*Fe$_2$As$_2$:Co systems. Figures 2(a–c) show the XRD patterns for bulk polycrystals of CaFe$_2$As$_2$:Co (a), SrFe$_2$As$_2$:Co (b), and BaFe$_2$As$_2$:Co (c), and their chemical compositions measured by XRF are shown for $AE$ = Ca below the XRD patterns. For $AE$ = Ca, we should note that an impurity CaFe$_4$As$_3$ phase was produced in our preliminal syntheses; however, it disappeared after further optimization of the synthesis process. As a result, the CaFe$_2$As$_2$:Co phase was obtained as the major phase at each reaction temperature between 800–1000 °C, as confirmed by the XRD patterns. However, single-phase samples have not yet been obtained. Reaction at 800 °C produced the CaFe$_2$As$_2$:Co phase, but coexisted with a small amount of FeAs impurity phase. With increasing reaction temperature up to 1000 °C, the amount of FeAs impurity increased. This increase was consistent with the XRF result showing that the Ca content in the synthesized samples decreased with increasing reaction temperature, which produced Ca-poor phases. Conversely, for the $AE$ = Sr and Ba cases, single-phase samples were obtained in the reactions at 900 °C. These results, along with our previous reports [13,43], indicate that the stability of these phases are in the order of BaFe$_2$As$_2$ > SrFe$_2$As$_2$ > CaFe$_2$As$_2$.

### 3.2. $T_s$ dependence of crystalline phases in thin films

Figure 3 shows XRD patterns of the $AE$Fe$_2$As$_2$:Co films ($AE$= Ca (a), Sr (b), and Ba (c)) on LSAT (001) grown at different $T_s$ between 600–1000 °C. We selected a growth rate of 3 Å/s and $t$ = 400 nm because these were typical values used when we reported epitaxial growth of SrFe$_2$As$_2$:Co in our early paper [10]. As will be shown later, this growth rate and $t$ are close to the optimum conditions for epitaxial growth of BaFe$_2$As$_2$:Co. At $T_s$ < 600 °C, no crystalline phase including $AE$Fe$_2$As$_2$:Co was detected by XRD. This result indicated that crystallization of all the $AE$Fe$_2$As$_2$:Co phases does not occur below 600 °C.

In the case of $AE$ = Ca (Fig. 3(a)), Ca-free impurity phases of $Fe_2As$, FeAs, $FeAs_2$, and Fe were mainly observed when $T_s$ was increased up to 650 °C, different to the results of the polycrystalline bulk syntheses, where only a small amount of FeAs impurity appeared and the $CaFe_2As_2$:Co phase was obtained as the major phase. We tried to reduce these Ca-poor impurities by increasing the nominal Ca content in the PLD target (i.e., the Ca content was increased up to Ca : Fe+Co : As = 2 : 2 : 2 in atomic ratio), but the impurity phase and content were not improved; therefore, the following experiments were performed using the stoichiometric target. The amount of Fe impurity, which was the dominant phase at $T_s$ = 700 °C, drastically increased as $T_s$ increased. These results suggested that severe off-stoichiometry of Ca occurred for thin films grown even at the low $T_s$ of 600–650 °C. We also performed XRF measurements for the thin films and found that 14 at.% of Ca remained in the films grown at $T_s$ = 650 °C (data not shown); this amount is smaller than the stoichiometric one (20 at.%) but cannot explain the XRD result that only the Ca-free phases were detected. This result suggests that Ca-containing amorphous phases had been formed. The Ca content decreased to 5 at.% and the As content to 7 at.% (stoichiometric content 40 at.%) at $T_s$ = 700 °C, which is consistent with the result that the Fe impurity increased with increasing $T_s$. Note that it is difficult to distinguish $Fe_2As$ and $CaFe_2As_2$:Co by XRD because their lattice parameters are similar ($a$ = 0.38942 and $c$ = 1.1746 nm for undoped $CaFe_2As_2$, $a$ = 0.38875 and $c$ = 1.1687 nm for $Ca(Fe_{0.97}Co_{0.03})_2As_2$ [21], and $a$ = 0.3632 and $2c$ = 1.1962 nm for $Fe_2As$). We confirmed that all the $2\theta$ angles of the diffraction peaks assigned to $Fe_2As$ in Fig. 2(a) were explained by the refined lattice parameters of $a$ = 0.3636 and $2c$ = 1.198 nm within a $2\theta$ error of 0.002 degrees. Only one peak, at 32.179 degrees cannot be assigned to $Fe_2As$ and is consistent with the 013 diffraction of $Ca(Fe_{0.97}Co_{0.03})_2As_2$ (calculated $2\theta = 32.512°$); the difference in $2\theta$ of 0.333 degrees can be attributed to the smaller content of Co in the obtained film. Therefore, we concluded that a small amount of $CaFe_2As_2$:Co phase formed in the film grown at $T_s = 650$ °C, but that the major phases were Ca-free impurities of $Fe_2As$, Fe, and $FeAs_2$. This is consistent with the electrical results to be discussed; i.e., superconductivity and resistivity anomaly, which are observed in $CaFe_2As_2$:Co single crystals [21], were not detected in these films. From the above results of Ca content decreasing with increasing $T_s$, we consider that the difficulty in growing a $CaFe_2As_2$:Co film in a vacuum chamber arises from the high vapor pressure of Ca.

On the other hand, in the case of $AE$ = Sr (Fig. 3(b)), a *c*-axis oriented film without an in-plane orientation was obtained at $T_s = 600$ °C along with impurity phases of FeAs observed at $2\theta = \sim35$ degrees and Fe at $2\theta = \sim65$ degrees, indicating that similar off-stoichiometry of the $AE$ element also occurs but also that more Sr is maintained in the films than Ca in the $CaFe_2As_2$:Co films. The *c*-axis orientation of the $SrFe_2As_2$:Co films was enhanced at $T_s = 700$ °C, where the epitaxial growth of the film is evidenced from the four clear peaks in the in-plane $\phi$ scan of the 200 diffraction (inset, Fig. 3(b)), but further increase in $T_s$ to 750 °C broadened the diffraction peaks, accompanied with an increase in the FeAs and Fe impurity amounts. At higher $T_s$ (800 °C), the diffraction peaks from the $SrFe_2As_2$ phase completely disappeared and only the FeAs and Fe phases were detected.

In the case of $AE$ = Ba (Fig. 3(c)), $BaFe_2As_2$:Co films without any preferential orientation were obtained at $T_s = 600$ °C, whereas a highly *c*-axis orientation was observed at higher $T_s = 700$–800 °C along with small extra peaks at $2\theta = 44$ and 65 degrees, which are respectively assigned to the 110 and 200 diffractions of Fe metal. We confirmed by cross-sectional transmission electron microscopy that the metal Fe impurity segregates only in the bulk region of the film, not at the interface and the film surface [44]. This result is clearly different from a report in ref. [45], where BaFe$_2$As$_2$:Co films were grown by PLD using a KrF excimer laser. The in-plane $\phi$ scans of the 200 diffractions of BaFe$_2$As$_2$:Co, shown on the right of each XRD pattern, confirmed their epitaxial growth on the LSAT substrates. At $T_s \geq 900$ °C, the preferential orientation changed from [001] to [110], and the 110 diffraction peak became dominant at $T_s = 1000$ °C. Even when $T_s$ was increased to 800 °C, the amount of the Fe impurity did not change, which implies that the BaFe$_2$As$_2$:Co phase is more stable at higher $T_s$ than SrFe$_2$As$_2$:Co, while a slight increase in Fe impurity was observed as $T_s$ increased from 800 to 1000 °C. These results indicate that the off-stoichiometry of Ba is much smaller than those of the $AE$ = Ca and Sr systems and that BaFe$_2$As$_2$:Co epitaxial films were obtained in the wider $T_s$ range of 700–900 °C compared with the $AE$ = Sr system, in which the $T_s$ window was very narrow (700–750 °C).

Figure 4 summarizes the above results of the $T_s$ dependence of the obtained crystalline phases. Off-stoichiometry of $AE$ in the obtained films becomes greater from $AE$ = Ba to Sr, and to Ca, leading to the observed narrower $T_s$ windows for $AE$ = Sr and in particular for Ca. Accordingly, in the $AE$ = Ca case, FeAs-related impurities were dominant in all the resulting films, while in the case of $AE$ = Sr the Fe and FeAs impurities, which became dominant at $T_s \geq 800$ °C, increased with $T_s$. In contrast, the BaFe$_2$As$_2$:Co phase was obtained even at high $T_s$ up to 1000 °C, but epitaxial growth was limited to the $T_s$ range of 700–900 °C. As explained in the former section,

*AE*Fe$_2$As$_2$:Co phases were obtained at 800–1000 °C for bulk polycrystalline samples reacted in evacuated silica-glass ampules. The narrow $T_s$ windows for thin film growth cannot be explained based on thermal equilibrium and would be attributed partly to deviation in the chemical compositions. The melting and boiling point temperatures ($T_m$ and $T_b$) and the vapor pressures at 527 °C ($P_v$) for the elemental *AE* metals are $T_m$ = 851 °C, $T_b$ = 1487 °C, and $P_v$ ~0.1 Pa for Ca, 774 °C, 1366 °C, and ~1 Pa for Sr, and 850 °C, 1537 °C, and ~6×10$^{-2}$ Pa for Ba [46,47]. Those of Sr and Ba may explain the narrower $T_s$ window of the *AE* = Sr phase compared with that of the *AE* = Ba phase, because $T_m$ and $T_b$ are lower and $P_v$ is higher for Sr. However, this cannot explain the results obtained for the *AE* = Ca phase because $T_m$ and $T_b$ are higher and $P_v$ is lower for Ca than those of Sr. In addition, first-principles calculations [48] provided very similar formation energies $\Delta E_f$ for *AE* = Ca, Sr, and Ba ($\Delta E_f$ = –14.42, –14.49, and –14.70 eV per molecule unit for Ca, Sr, and Ba, respectively, for the formation reactions of *AE*As + Fe$_2$As → *AE*Fe$_2$As$_2$) and cannot explain the difficulty in obtaining the CaFe$_2$As$_2$ phase. A possible reason may be a slow reaction between Ca and other intermediate reagents due to the high stability of the elemental Ca metal, as speculated from its $T_m$, $T_b$, and $P_v$. Further study is required to elucidate its origin.

Due to the narrow $T_s$ window (700–750 °C) for SrFe$_2$As$_2$:Co epitaxial growth, a very limited range of process parameters remains to improve the film quality, and this could be a reason why the SrFe$_2$As$_2$:Co films exhibited broad superconducting transition ($\Delta T_c = T_c^{\text{onset}} – T_c^{\text{offset}}$ = 6 K) [10] and a weak-link behavior [49]. Conversely, the wider $T_s$ window (700–900 °C) for growing BaFe$_2$As$_2$:Co epitaxial films provides much larger flexibility to improve the film quality by optimizing other growth parameters. In addition, the BaFe$_2$As$_2$:Co phase is more stable in moisture-containing ambient air than SrFe$_2$As$_2$:Co [13]. BaFe$_2$As$_2$:Co films have therefore been most-widely studied among the 122-type *AE*Fe$_2$As$_2$:Co compounds and exhibited the best performance to date ($\Delta T_c$ = 1 K, $J_c^{trans}$ > 1 MA/cm$^2$) among the iron-based superconductor films [14,22,23].

Note that rare-earth (*RE*) doped CaFe$_2$As$_2$ appears to be another promising candidate for a high-$T_c$ superconductor because of its high $T_c^{onset}$, > 40 K [50,51], which is the highest value among the 122-type *AE*Fe$_2$As$_2$ compounds, although there is still controversy over the origin of this high $T_c$ [52]. However, as shown in this section, a CaFe$_2$As$_2$ film has not yet been produced by PLD primarily due to the large off-stoichiometry of Ca. Therefore, we recently fabricated *RE*-doped BaFe$_2$As$_2$, which has not been obtained in the bulk form, thin films on MgO single-crystal substrates because we expected that a high $T_c$, similar to that of the *RE*-doped CaFe$_2$As$_2$, would be obtained if *RE*-doping could be realized for BaFe$_2$As$_2$ [53]. As a result, the non-equilibrium PLD film growth process effectively stabilized the metastable doping of La at Ba sites. But, unlike the *RE*-doped CaFe$_2$As$_2$, the obtained maximum $T_c$ were ~22 K for (Ba,La)Fe$_2$As$_2$, close to those of the BaFe$_2$As$_2$:Co films and (Sr,La)Fe$_2$As$_2$ polycrystals synthesized under high pressures at 2–3 GPa [54].

### 3.3. Variation of surface morphology with $T_s$ for BaFe$_2$As$_2$:Co films

Next, we discuss the surface morphology of 400-nm-thick BaFe$_2$As$_2$:Co films on LSAT (001) grown at different $T_s$ (600–1000 °C) and a fixed growth rate of 3 Å/s. Figure 5 shows AFM images of the BaFe$_2$As$_2$:Co films. Polycrystalline films were grown at $T_s$ = 600 °C and had randomly-oriented planar-shaped grain structures, which is consistent with the XRD result in Fig. 3(c). The more isotropic island growth began at $T_s$ = 700 °C. With further increases in $T_s$, the domain size increased and the growth mode changed from a three-dimensional island mode to a step-flow mode at $T_s$ = 800–850 °C. Although a step-and-terrace structure with a step height of 1.3 nm, which is comparable to the *c*-axis length of BaFe$_2$As$_2$:Co, was seen in the flat region [13], some sub-μm-sized void structures coexisted, and particulates with diameters of ~0.1 μm were observed at the film surfaces. Increase in $T_s$ to > 850 °C regenerated the island structures and the surface morphology became extremely rough. This trend is explained well by the XRD results, in which the crystallographic orientation was observed to change from epitaxial to 110-preferentially oriented polycrystals. Consequently, at $T_s$ = 1000 °C, completely misoriented grains and granular growth (lateral grain size: ca. 500 nm) were clearly observed. Therefore, based on the XRD and AFM results, we concluded that the optimum $T_s$ for BaFe$_2$As$_2$:Co epitaxial films is 800–850 °C under the fixed growth rate and thickness used here.

### 3.4. Changes in crystallinity and $J_c^{\text{trans}}$ of BaFe$_2$As$_2$:Co films with $T_s$

Figure 6(a) summarizes the $T_s$ dependence of crystallinity for the BaFe$_2$As$_2$:Co films. The growth rate and thickness were kept at constant values of 3 Å/s and 400 nm, respectively. The crystallinity was characterized as the full width at half maximum (FWHM) values of the 002 rocking curves ($\Delta\omega$) and the 200 rocking curves ($\Delta\phi$) obtained by out-of-plane $2\theta$-fixed $\omega$ scans and in-plane $2\theta_\chi$-fixed $\phi$ scans, respectively. With increasing $T_s$ from 700 °C, both $\Delta\omega$ and $\Delta\phi$ were improved, leading to minimum values (FWHMs of $\Delta\omega$ and $\Delta\phi$ = 0.5 degrees) at $T_s$ = 830 °C. It is noteworthy that these FWHMs are comparable with or even smaller than those of BaFe$_2$As$_2$:Co films grown using buffer layer techniques [14,37,55]. Conversely, at $T_s \geq$ 900 °C, where the crystallographic orientation changed (as shown in Fig. 3(c)), slight increases in FWHMs were observed for both $\Delta\omega$ and $\Delta\phi$. Therefore, the optimum $T_s$ range is 800–850 °C, which is consistent with the above AFM results. In the case of the $AE$ = Sr films grown at $T_s$ = 700 °C [10], the FWHMs of both $\Delta\omega$ and $\Delta\phi$ were ~1 degree. These values are slightly smaller than those of the BaFe$_2$As$_2$:Co films grown at the same $T_s$. On the other hand, we can employ higher $T_s$ for BaFe$_2$As$_2$:Co owing to the wider $T_s$ window, shown in Fig. 4, and higher crystallinity was achieved in these BaFe$_2$As$_2$:Co epitaxial films than in the SrFe$_2$As$_2$:Co films by optimizing $T_s$.

Accompanying the improvement of crystallinity, $J_c^{trans}$ at 4 K increased from 0.1 to $\geq$ 1 MA/cm$^2$ (Figure 6(b)), while at $\geq$ 900 °C a drastic decrease in $J_c^{trans}$ was observed depending on the degradation of crystalline orientation, which led to a weak link behavior. These results indicate that the optimum $T_s$ is 800–850 °C, which is consistent with the above high-resolution XRD result, while the observed angular dependence of $J_c^{trans}$ [56] clarified that the films grown at the optimum condition have strong $c$-axis oriented pinning centers. Therefore, change in $T_s$ from 700 to 850 °C induces improvement of the crystallinity, and, at the same time, introduces defects working as effective pinning centers.

Here, we would like to note that the same $T_s$ dependence of crystallinity and $J_c^{trans}$ were obtained for BaFe$_2$As$_2$:Co films on MgO (001) single crystals (see Supplementary Information) despite the in-plane lattice mismatch between BaFe$_2$As$_2$:Co and MgO (+6 %) being larger than that in the case of LSAT (001) (–2 %). In addition, not only $J_c^{trans}$ but also $\Delta T_c$ were improved from 3 to 1 K [22]. These results indicate that the optimum $T_s$ condition ($T_s$ = 800–850 °C) obtained from the XRD and AFM observations is also appropriate for obtaining BaFe$_2$As$_2$:Co films with high crystallinity and high $J_c^{\text{trans}}$.

### 3.5. Relationship between growth rate and crystallinity of BaFe$_2$As$_2$:Co films

In the previous sections, we refined the optimum $T_s$ for BaFe$_2$As$_2$:Co epitaxial films to 800–850 °C. However, we noticed that the growth rate also significantly affected the film growth mode. Figure 7 summarizes the growth rate dependence of the crystallite orientations and surface morphologies of 400-nm-thick BaFe$_2$As$_2$:Co films on LSAT (001) grown at $T_s$ = 800 °C. Figure 7(a) shows the XRD patterns of the films grown at different growth rates of 1.6–4.3 Å/s, which were controlled by changing the fluence of the Nd:YAG laser from 1.7 to 4.3 J/cm$^2$/pulse. At a low growth rate of 1.6 Å/s, preferentially [110] orientation was observed, and epitaxial growth was observed at growth rates ≥ 2.8 Å/s. In contrast, higher growth rates ≥ 3.7 Å/s led to generation of diffraction peaks other than the 110 peak of BaFe$_2$As$_2$:Co, which indicated the coexistence of epitaxial and polycrystalline growth. Figure 7(b) summarizes the AFM images of the resulting films. Films grown at the lowest growth rate had flat surface regions and grains oriented in different directions. In contrast, the films grown at higher rates of 2.8 and 3.3 Å/s showed flat surfaces as shown in Fig. 5(d). It was clearly evident that the surface became rougher with further increases of growth rate ≥ 3.7 Å/s because of the resulting three dimensional growth and the generation of polycrystalline phases. These results substantiate that precise control of the growth rate to 2.8–3.3 Å/s is an important factor for the epitaxial growth of high-quality BaFe$_2$As$_2$:Co films.

### 3.6. Thickness dependence of crystallite orientation and $J_c^{\text{trans}}$ in BaFe$_2$As$_2$:Co films

For application to superconducting tapes with high-density current drivability, thick films are required. Thus, we investigated film thickness dependence for BaFe$_2$As$_2$:Co films grown on LSAT (001) under optimum growth conditions ($T_s$ = 850 °C and growth rate = 2.8 Å/s). $t$ was varied from 110 to 1080 nm by changing the deposition time. Figure 8(a) shows XRD patterns of the BaFe$_2$As$_2$:Co films with different $t$. Clear $c$-axis orientation was observed up to $t$ = 570 nm, but other diffraction peaks attributed to polycrystalline BaFe$_2$As$_2$:Co were observed at $t$ = 1080 nm. This result indicates that the coherent epitaxial growth on LSAT substrates proceeds up to $t$ = 570 nm but new nucleation occurs at the growing surface to produce randomly-oriented grains between 570 and 1080 nm. In the case of an Fe metal buffer layer [55], similar coherent growth occurs only when $t$ < 90 nm, and a 110 diffraction peak begins to be observed on thicker films. This implies that the factors affecting the growth mode, e.g., nucleation and strain, are different between our direct deposition case and the latter buffer layer case.

We confirmed that $T_c$ and $\Delta T_c$ were independent of $t$. Here, we examined the effect of $t$ on $J_c$. Figure 8(b) summarizes the $J_c^{\text{trans}}$ at 4 K of BaFe$_2$As$_2$:Co films with different $t$. Although the $J_c^{\text{trans}}$ values exhibit some scattering, they are almost constant at ~2 MA/cm$^2$ up to 480 nm. For larger $t$, $J_c^{\text{trans}}$ decreased to ~1 MA/cm$^2$ as $t$ increased. A similar result (i.e., weak thickness dependence of $J_c^{\text{trans}}$) has been reported in YBCO films that contain BaZrO$_3$ nanorod structures; with the explanation that their strong $c$-axis pinning centers are effective in reducing the thickness dependence of $J_c^{\text{trans}}$ [56]. In the case of the present BaFe$_2$As$_2$:Co films, naturally-formed defects in the films behave as strong $c$-axis pinning centers with large pinning force $F_p$ > 30 GNm$^{-3}$ at ~12 T [57]. Therefore, it is speculated that the $J_c$–$t$ relation for $t \leq 480$ nm originates from such *c*-axis pinnings. On the other hand, possible factors contributing to the decreased $J_c^{trans}$ for thick films are the weakening of the in-plane and out-of-plane texture, poorer crystalline qualities, and/or weakening of the pinning force due to change in defect structures with *t*. The first would be important for the 1080-nm thick films and the others may have affected the 570-nm thick films. To compare the pinning properties in a magnetic field, we measured *H* dependence of $J_c^{mag}$ ($J_c^{mag}(H)$) for epitaxial films with similar crystallinity but with different thicknesses of *t* = 110, 350, and 570 nm. Figure 8(c) shows $J_c^{mag}(H)$ curves at 3.5 K up to $\mu_0H = 3$ T. The slope coefficient of $|\partial J_c^{mag}(H) / \partial H|$ is larger for the thicker film with *t* = 570 nm, presumably due to weak flux pinning, which would cause the decrease in $J_c^{trans}$ with increasing *t*. Further experiments, such as microstructure analyses and angular dependences of $J_c^{trans}$, are necessary to fully understand the $J_c^{trans}$–*t* relation. However, we concluded that *t* = 100–500 nm are appropriate in terms of optimized $J_c$ for BaFe$_2$As$_2$:Co films because both $J_c^{trans}$ and $J_c^{mag}$ begin to decrease at *t* = 570 nm, although coherent epitaxial growth still proceeds.

### 3.7. Application of the optimum growth condition to Ba(Fe$_{1-x}$Co$_x$)$_2$As$_2$ films with various *x* values

In the previous sections, we found optimum growth conditions for the BaFe$_2$As$_2$:Co epitaxial film with *x* = 0.08 to be $T_s$ = 800–850 °C, growth rate between 2.8 and 3.3 Å/s, and *t* = 100–500 nm. Finally, we applied these optimum conditions to BaFe$_2$As$_2$:Co epitaxial films with various doping concentrations (*x*). Figure 9(a) shows the XRD patterns of 300-nm-thick BaFe$_2$As$_2$:Co epitaxial films with nominal *x* values of 0.03–0.1, which were controlled by changing *x* in the PLD targets. As seen in Fig. 9(a), all the films were preferentially oriented to the *c*-axis and each 00*l* diffraction peak shifted to higher angles with increasing $x$, indicating reduction in $c$-axis length. This shrinkage of the $c$-axis with increasing $x$ is consistent with the result for $BaFe_2As_2$:Co single crystals [42]. Figure 9(b) shows $\rho$–$T$ curves scaled by $\rho$ at 300 K ($\rho$ (300 K)). A resistivity anomaly, associated with structural and magnetic phase transitions [58], was observed at $x = 0.03$ and 0.05. As $x$ increased, the resistivity anomaly temperature ($T_{anom}$), at which $|d\rho/dT|$ reaches maximum values, shifted to lower $T$, and the anomaly was not observed at $x = 0.075$. We fitted the $\rho$–$T$ curves, in a normal metal state at $T \geq$ 150 K, using a conventionally-used power-law behavior, $\rho = A + BT^n$. The inset to Fig. 9(b) shows the fitting results, and Fig. 9(c) compares the $x$ dependence of the $n$ values with those reported for $BaFe_2As_2$:Co single crystals [59,60], for which the $n$ value has been shown to decrease with $x$ and take a minimum $n = \sim 1$ around $x = 0.08$ before increasing again with increasing $x$. The $n$ values of the $BaFe_2As_2$:Co epitaxial films also exhibited the same behavior as those reported for single crystals; $n = 1$, that of the optimally doped $BaFe_2As_2$:Co single crystals [59], was obtained at $x = 0.075$. These results indicate that the nominal $x = 0.03$–0.1 of the $BaFe_2As_2$:Co films are almost identical to $x$ values for single crystals measured with an energy dispersive x-ray spectrometer [61], i.e., $x$ is successfully controlled in the films by changing the nominal $x$ in the PLD targets [62]. In addition, $T_c$ (shown in the bottom figure of Fig. 9(b)) exhibits a systematic variation similar to the $n$ value (Fig. 9(d)), i.e., $T_c$ increases with $x$ at $x < 0.075$ and turns to decrease at higher $x$, which is the same behavior as that reported for single crystals [61]. As for the transition behavior, $\Delta T_c$ were slightly broader in the low doping concentration region ($x = 0.03$ and 0.05) than those observed in the higher region. This broadening of $\Delta T_c$ for low $x$ was also observed for single crystals [61]. A maximum $T_c$ of 25.5 K was obtained for the optimal $x = 0.075$; this $T_c$ is comparable to the maximum $T_c$ ever reported for BaFe$_2$As$_2$:Co single crystals [63] and thin films [15].

## 4. Summary

We examined the effects of growth parameters on stability, structure, and physical properties of $AE$Fe$_2$As$_2$:Co ($AE$ = Ca, Sr, and Ba) thin films prepared by PLD using a second harmonic Nd:YAG laser and proposed optimum conditions for high-quality BaFe$_2$As$_2$:Co epitaxial films with high $T_c$ and $J_c$.

It was found that substrate temperature $T_s$ is a critical parameter to produce the $AE$Fe$_2$As$_2$:Co crystalline phase. No crystalline phase was detected by XRD at $T_s$ < 600 °C. It is very difficult to grow crystalline phase of $AE$ = Ca even for higher $T_s$; it decomposes to Ca-free impurities Fe$_2$As, FeAs$_2$, FeAs, and Fe due to selective evaporation of Ca. The $AE$ = Sr phase was obtained in the $T_s$ range 600–750 °C but decomposed to Fe and FeAs at higher temperatures. As a consequence, epitaxial growth for the $AE$ = Sr phase is limited to the narrow $T_s$ range of 700–750 °C. In contrast, the off-stoichiometry of Ba is much smaller than for Ca and Sr, even for the high-$T_s$ growth, and thus the BaFe$_2$As$_2$:Co phase was obtained in a wider $T_s$ range up to 1000 °C.

For BaFe$_2$As$_2$:Co film growth, not only $T_s$ but also the growth rate affect the surface morphology, crystallographic orientation, and crystallinity. Epitaxial growth with a two-dimensional growth mode was observed only in the $T_s$ range of 800–850 °C at a growth rate between 2.8–3.3 Å/s, leading to high crystallinity and a maximum $J_c^{trans}$ of 4 MA/cm$^2$ at 4 K. In addition, it was found that film thicknesses of 100–500 nm are most suitable in order to obtain BaFe$_2$As$_2$:Co epitaxial films with high $J_c$, > 1 MA/cm$^2$. The optimum growth conditions ($T_s$ = 800–850 °C, growth rate = 2.8–3.3 Å/s, and $t$ =

100–500 nm) were applicable to the entire $x$ range for $Ba(Fe_{1-x}Co_x)_2As_2$ films, and yielded a high $T_c^{onset}$ of 25.5 K for the nominal $x = 0.075$.

## Acknowledgment

This work was supported by the Japan Society for the Promotion of Science (JSPS), Japan, through the "Funding Program for World-Leading Innovative R&D on Science and Technology (FIRST Program)".

## Figure captions

**Figure 1.** (a) The pulsed laser deposition system used in this study. It consists of a preparation chamber, a growth chamber, and a Nd:YAG laser. Photographs of a target disk and the inside of the growth chamber during thin film growth are also shown. (b) Substrate carrier. A platinum sheet is placed on the center of the carrier to make a good thermal contact with the substrate. (c) The single-crystal substrate clamped to the substrate carrier by the top cover. (d) Side-view cross-section of the substrate carrier setup.

**Figure 2.** (a) XRD patterns for $Ca(Fe_{0.97}Co_{0.03})_2As_2$ bulk polycrystals synthesized at reaction temperatures of 800 °C (shown in bottom panel), 900 °C (middle), and 1000 °C (top). Diffraction peaks of the FeAs impurity phase are indicated by pink vertical bars. Weight fractions of the obtained crystalline phases, estimated by Rietveld analyses using TOPAS code, are shown at the top-left of each figure. The figure shown below the XRD patterns represents the atomic ratio percentages of Ca, (Fe+Co), and As of the bulk polycrystals as measured by XRF analyses. (b) and (c) show the XRD patterns of bulk polycrystals for $Sr(Fe_{0.9}Co_{0.1})_2As_2$ (b) and $Ba(Fe_{0.92}Co_{0.08})_2As_2$ (c), synthesized at 900 °C. Peak positions of each $AE$Fe$_2$As$_2$ phase are denoted by the black vertical bars at the bottom of each XRD pattern. Miller indexes are noted above the corresponding diffraction peaks.

**Figure 3.** XRD patterns of $AE$Fe$_2$As$_2$:Co films ($AE$ = Ca (a), Sr (b), and Ba (c)) grown at $T_s$ = 600–1000 °C. The $T_s$ for each XRD pattern is shown on the left. The diffraction peaks of $AE$Fe$_2$As$_2$:Co phases are indicated by the vertical bars, and those of Fe, FeAs,

Fe$_2$As, and FeAs$_2$ impurities by circles, triangles, squares, and diamonds, respectively. The right-side inset figures are in-plane $\phi$-patterns of 200 diffractions of the $AE$Fe$_2$As$_2$:Co phases.

**Figure 4.** Summary of the relationship between $T_s$ and crystalline phases in thin films.

**Figure 5.** AFM images of 2 μm × 2 μm surface areas of BaFe$_2$As$_2$:Co films grown at $T_s$ from (a) 600 to (g) 1000 °C. Horizontal scale-bar below each image shows the height scale.

**Figure 6.** $T_s$ dependence of crystallinity and $J_c^{\text{trans}}$ for BaFe$_2$As$_2$:Co films. (a) Change in FWHM values of $\Delta\omega$ of 002 rocking curves (closed triangles) and $\Delta\phi$ of 200 rocking curves (open circles) with $T_s$. (b) Relationship between $J_c^{\text{trans}}$ at 4 K and $T_s$.

**Figure 7.** (a) XRD patterns and (b) AFM images of 2 μm × 2 μm areas of BaFe$_2$As$_2$:Co films grown at $T_s$ = 800 °C with different growth rates of 1.6–4.3 Å/s. Growth rate is shown on the right of each XRD pattern and on the upper left of each AFM image. The asterisks in (a) indicate Fe impurity diffraction peaks. Horizontal scale-bar below each image indicates the height scale.

**Figure 8.** Film thickness ($t$) dependence of crystallite orientations and $J_c$ for BaFe$_2$As$_2$:Co films grown under optimum conditions. (a) XRD patterns of BaFe$_2$As$_2$:Co films with different $t$ of 110–1080 nm. Film thicknesses are shown on the right. The asterisks indicate Fe impurity diffraction peaks. (b) $t$ dependence of $J_c^{\text{trans}}$ at 4 K in a self-field condition for BaFe$_2$As$_2$:Co films. (c) Magnetic field dependence of $J_c^{\text{mag}}$

($J_c^{mag}(H)$) at 3.5 K for $BaFe_2As_2$:Co films with $t$ of 110, 370, and 570 nm.

**Figure 9.** Change in structural and electrical properties of $BaFe_2As_2$:Co epitaxial films with different nominal doping concentrations ($x$ in $Ba(Fe_{1-x}Co_x)_2As_2$). (a) XRD patterns for films with $x$ = 0.03–0.1. The asterisks indicate Fe impurity diffraction peaks. The dotted lines show peak positions of the 00$l$ diffraction of the $x$ = 0.03 film. (b) $\rho$–$T$ curves normalized by $\rho$ at 300 K, $\rho$ / $\rho$(300 K), for films with $x$ = 0.03–0.1. $T_{anom}$ are indicated by black arrows. Inset figure shows the fitting results using an empirical power-law relation $\rho = A + BT^n$ at $T \geq 150$ K. Bottom panel is a magnified view around $T_c$. (c) $x$ dependence of $n$ values estimated from the $\rho$–$T$ curves for the epitaxial films. For comparison, those reported for $BaFe_2As_2$:Co single crystals are also shown in the figure [59,60]. The closed and open symbols represent data for epitaxial films and single crystals (SC), respectively. (d) Relationship between $T_c$ ($T_c^{onset}$: circles, $T_c^{offset}$: triangles) and $x$. Those reported for single crystals are also shown for comparison [61].

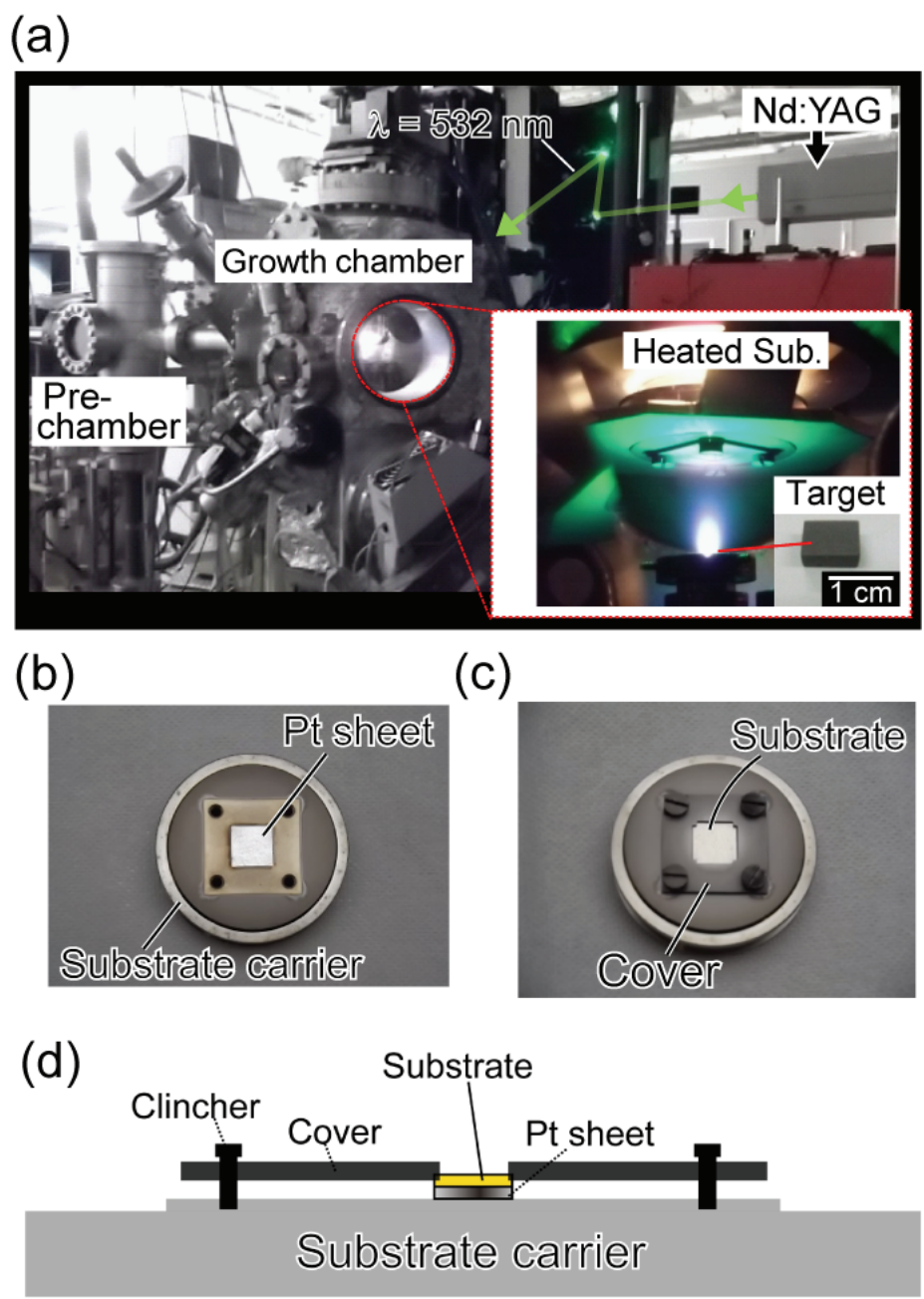

Figure 1

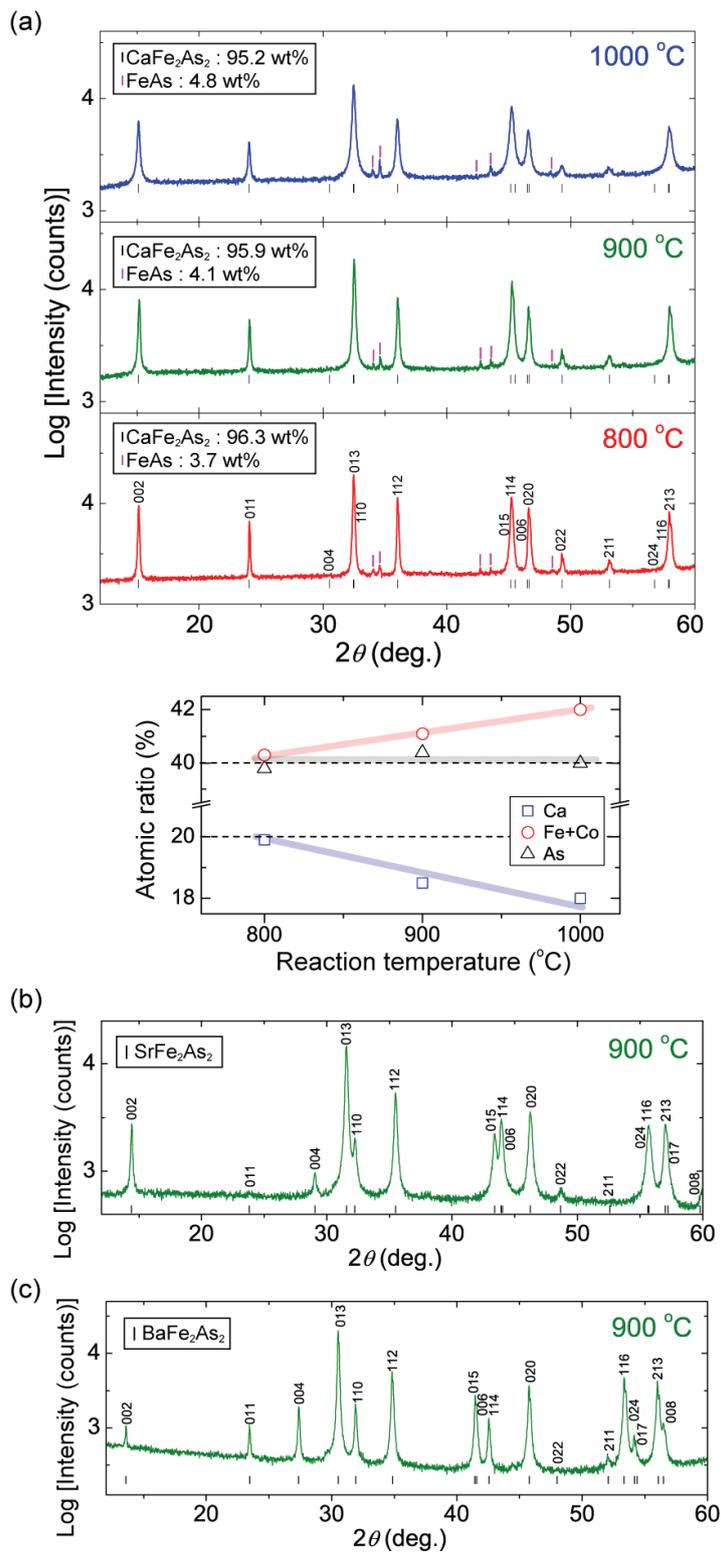

Figure 2

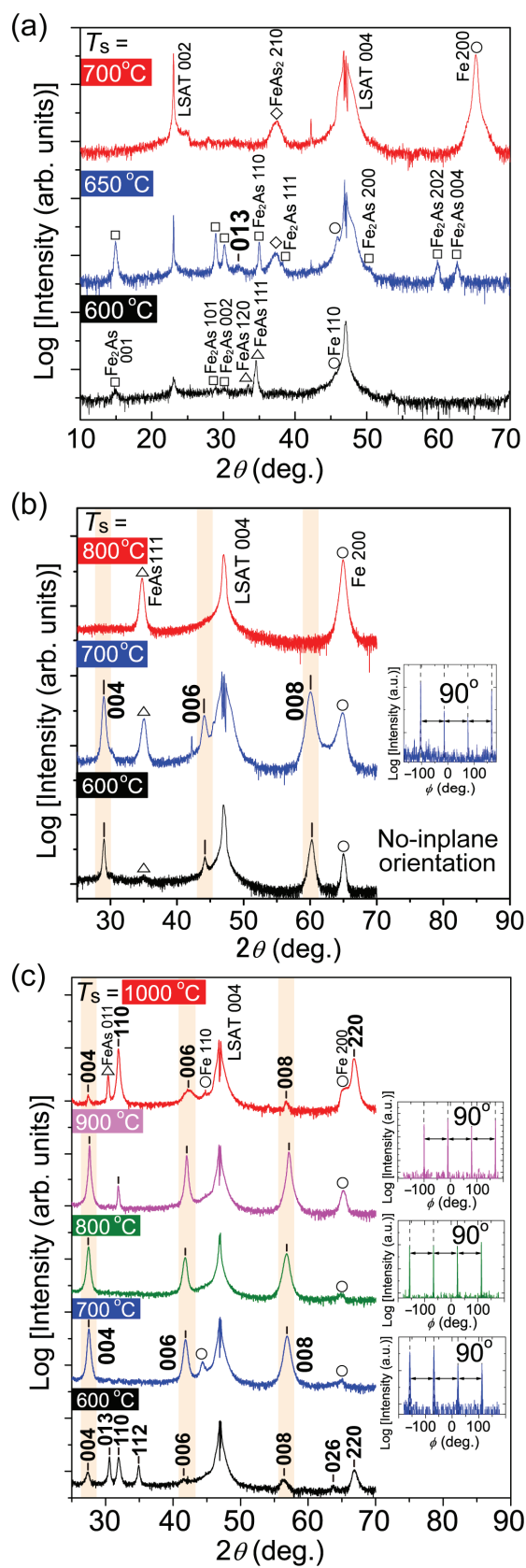

Figure 3

$T_S$  600 °C ▬▬ 650 °C ▬▬ 700 °C ▬ 750 °C ▬ 900 °C ▬▶ 1000 °C

| | 600 °C – 650 °C | 650 °C – 700 °C | 700 °C – 750 °C | 750 °C – 900 °C | 900 °C – 1000 °C |
|---|---|---|---|---|---|
| $CaFe_2As_2$:Co | $Fe_2As$, FeAs, Fe | Polycrystalline, Fe, $Fe_2As$, $FeAs_2$ | Fe, $FeAs_2$ | —— | —— |
| $SrFe_2As_2$:Co | *c*-axis oriented, Fe,FeAs | | **Epitaxial**, Fe, FeAs | Fe, FeAs | —— |
| $BaFe_2As_2$:Co | Polycrystalline, Fe | | **Epitaxial**, Fe | | 110-oriented polycrystalline, Fe |

Figure 4

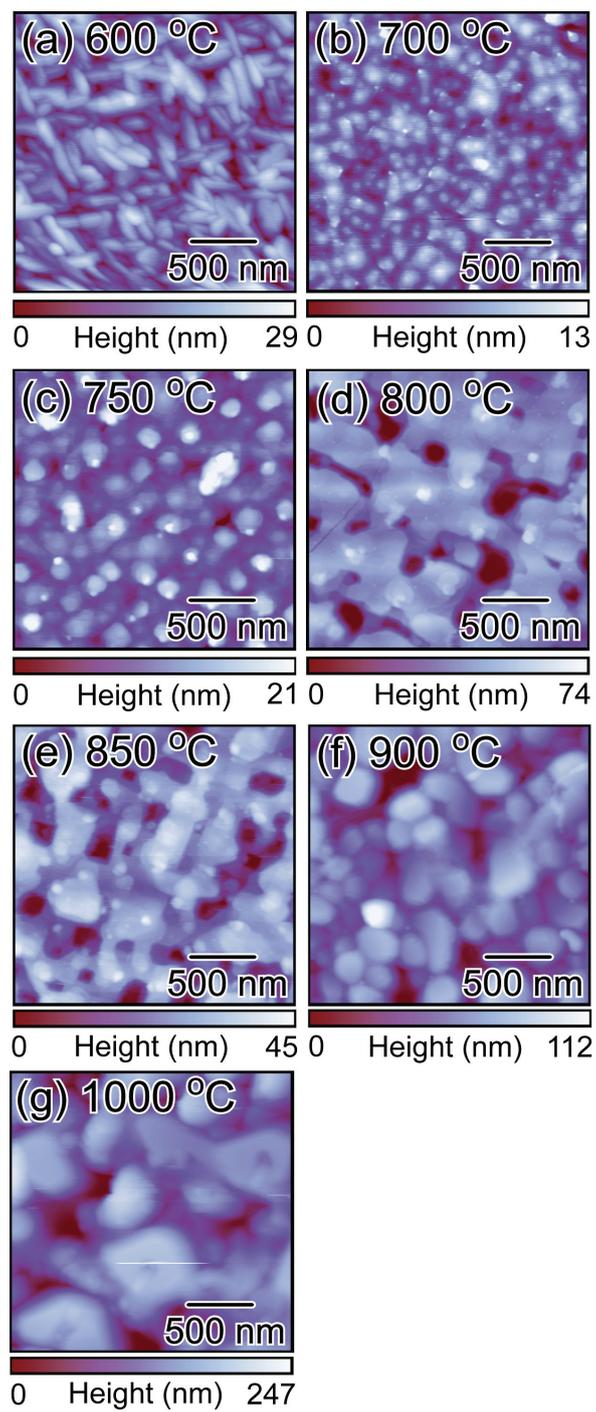

Figure 5

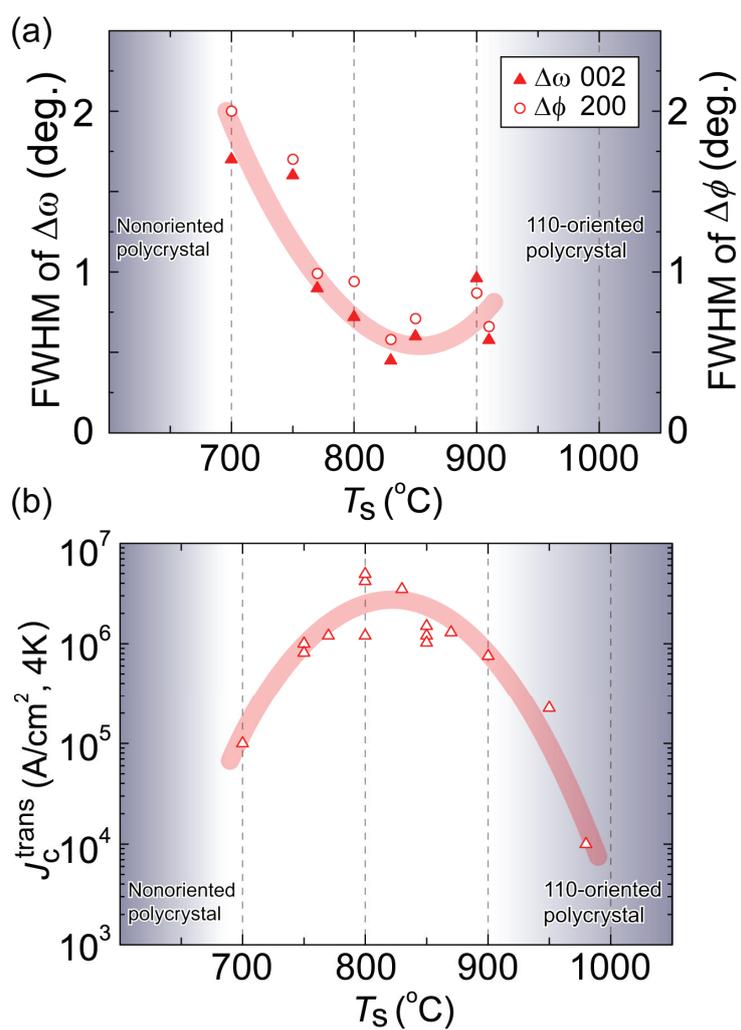

Figure 6

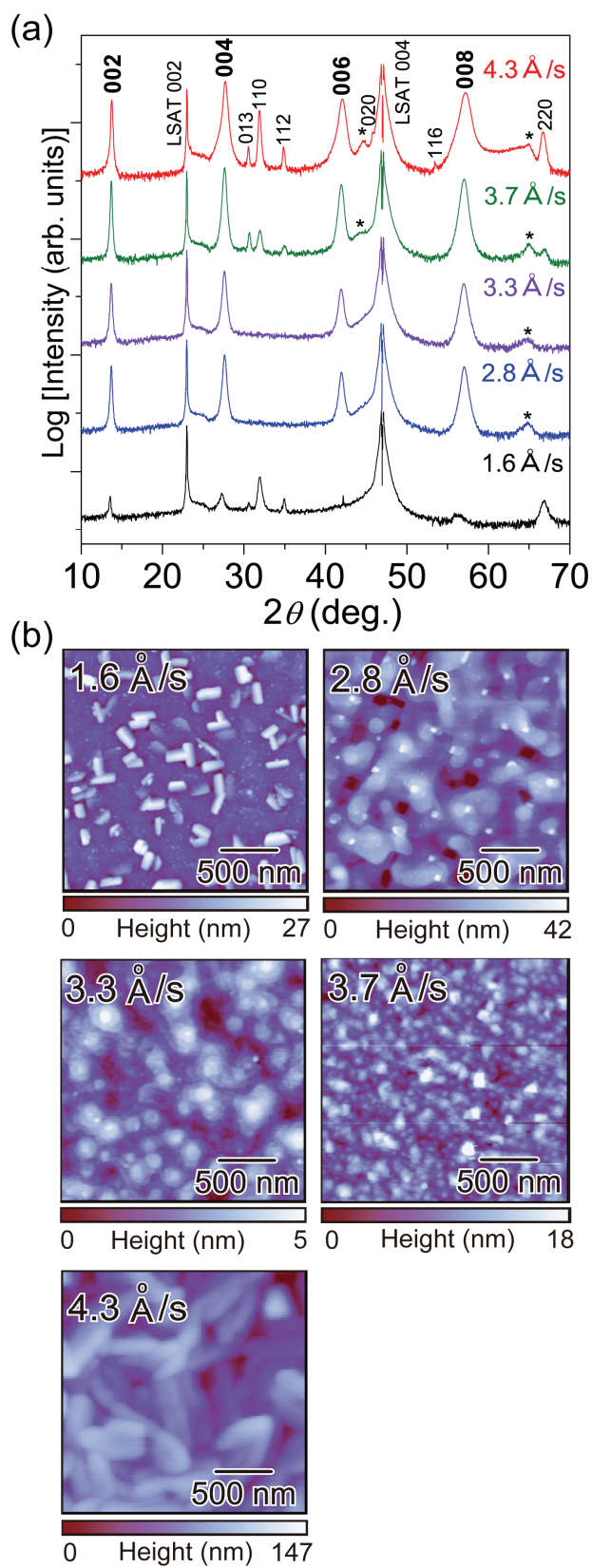

Figure 7

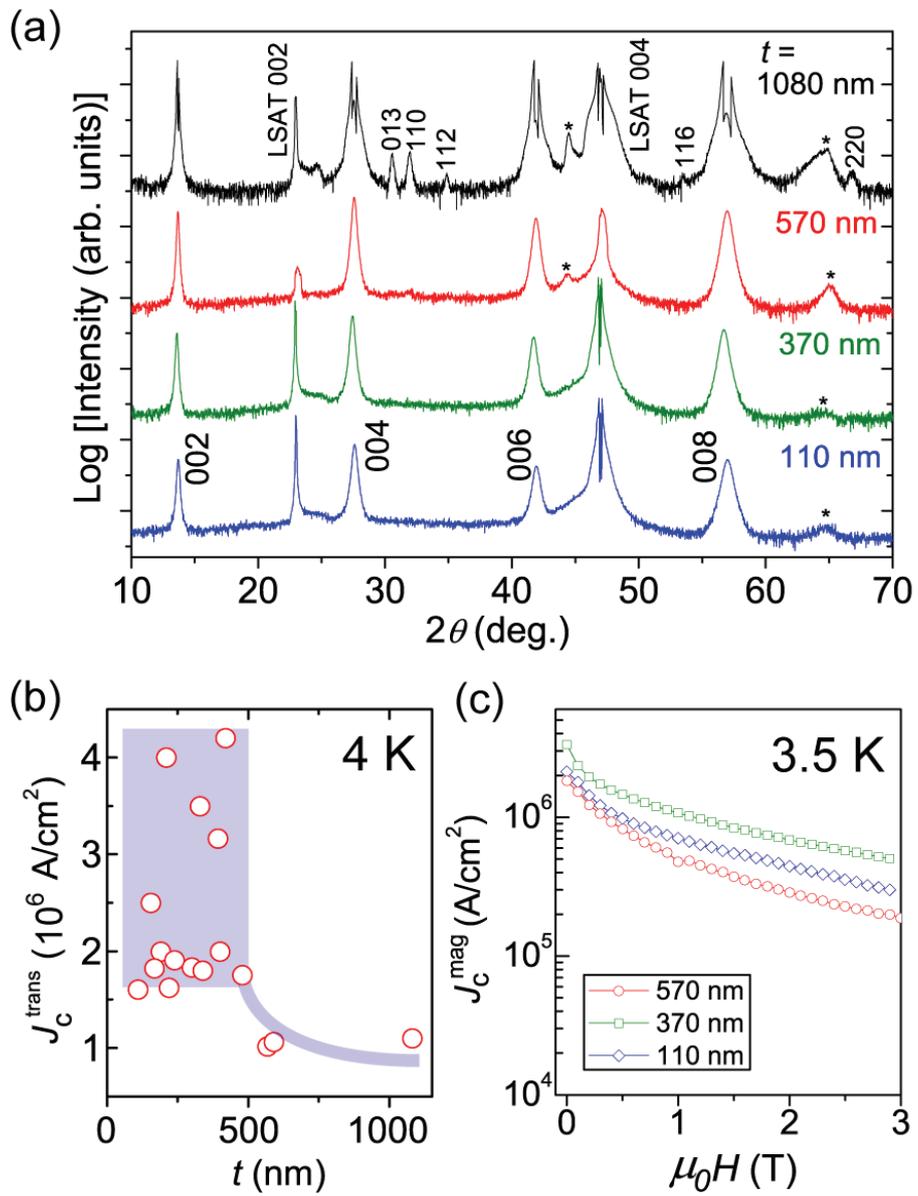

Figure 8

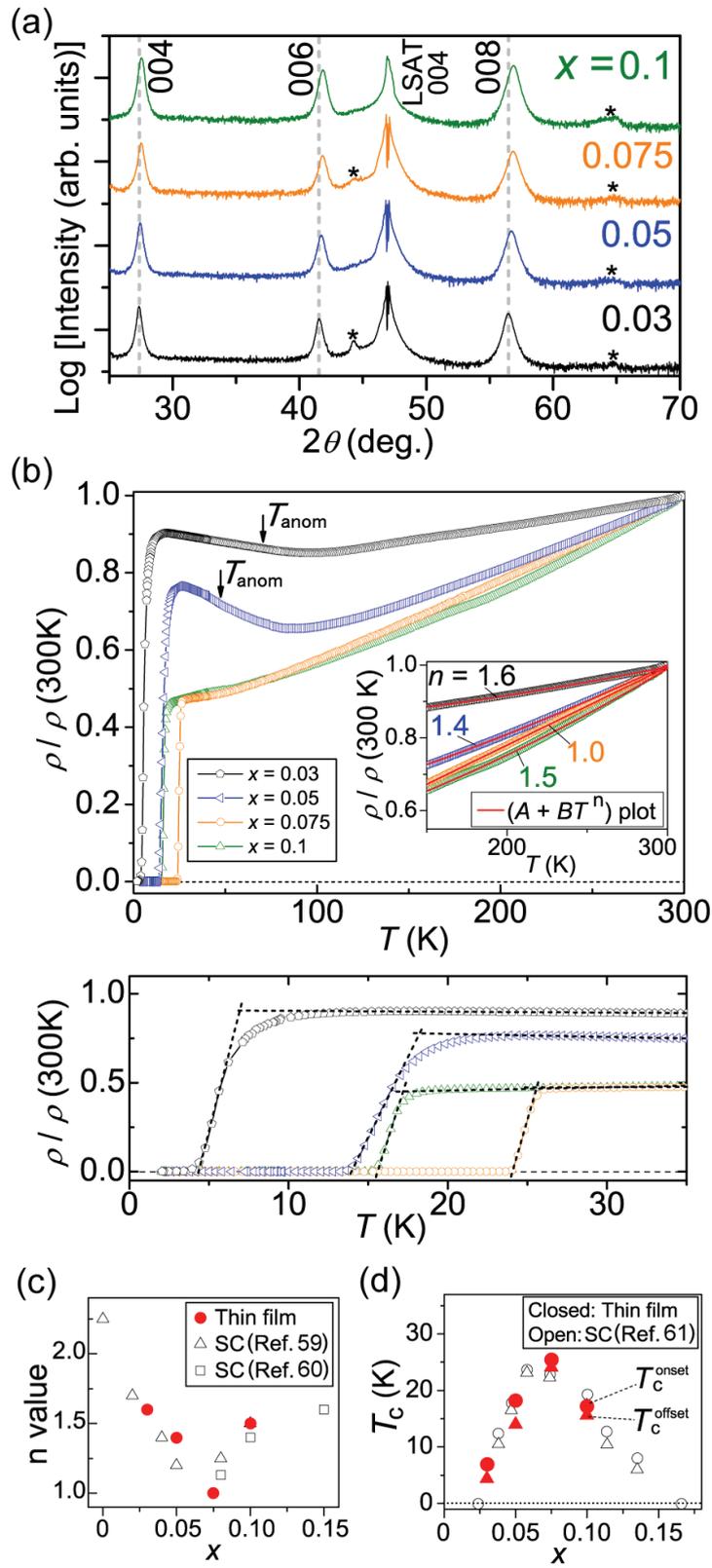

Figure 9